\documentclass[twocolumn]{aastex631}
\usepackage{textcomp}
\usepackage{amssymb}
\usepackage{rotating}
\usepackage{threeparttable}

\newcommand{\fermi}{\textit{Fermi}}
\newcommand{\gr}{$\gamma$-ray}
\newcommand{\hawc}{J0631+107}
\newcommand{\lasso}{J0631+1040}
\newcommand{\psr}{J0631$+$1036}

\shorttitle{A TeV halo associated with PSR J0631+1036?}
\shortauthors{Zheng et al.}

\begin{document}

\title{3HWC J0631+107/LHAASO J0631+1040: a TeV halo powered by the pulsar J0631+1036?}


\author{Dong Zheng}
\affiliation{Department of Astronomy, School of Physics and Astronomy, Yunnan University, Kunming 650091, China; wangzx20@ynu.edu.cn}

\author[0000-0003-1984-3852]{Zhongxiang Wang}
\affiliation{Department of Astronomy, School of Physics and Astronomy,
Yunnan University, Kunming 650091, China; wangzx20@ynu.edu.cn}
\affiliation{Key Laboratory for Research in Galaxies and Cosmology,
Shanghai Astronomical Observatory, Chinese Academy of Sciences, 
80 Nandan Road, Shanghai 200030, China}

\author{Yi Xing}
\affiliation{Key Laboratory for Research in Galaxies and Cosmology,
Shanghai Astronomical Observatory, Chinese Academy of Sciences,
80 Nandan Road, Shanghai 200030, China}

\begin{abstract}
	PSR~J0631+1036 is a middle-aged pulsar with properties similar to those
	of the nearby Geminga pulsar. It is bright in $\gamma$-rays, and has
	been noted as the only source possibly associated with the TeV source
	3HWC J0631+107 (also the LHAASO J0631+1040). For understanding the 
	nature of the TeV source, we analyze the GeV $\gamma$-ray data 
	obtained with the Large Area Telescope (LAT) onboard 
	{\it the Fermi Gamma-ray Space Telescope} for the source region.  
	We are able to remove the pulsar's emission from the region from timing 
	analysis, and find that the region is rather clean without 
	possible GeV $\gamma$-ray emission present as the counterpart to
	the TeV source. By comparing this pulsar to Geminga and considering
	the spectral feature of the TeV source, we argue that
	it is likely the TeV halo powered by the pulsar.
\end{abstract}


\section{Introduction}
PSR~\psr, discovered by \citet{zep+96}, is a middle-aged one having
spin period $P\simeq 0.288$\,s,
characteristic age $\tau_c\simeq$43.6\,kyr, and spin-down luminosity 
$\dot{E}\simeq 1.7\times 10^{35}$\,erg\,s$^{-1}$. Based on the new 
electron-density model for the Galaxy \citep{ymw17}, its distance 
is $D\simeq 2.1$\,kpc given in the Australia Telescope National Facility 
Pulsar Catalogue \citep{man+05}. In X-rays, observational studies have 
not detected the pulsar down 
to $\simeq 4.9\times 10^{30} (D/2.1)^{2}$ erg\,s$^{-1}$ (in 0.5--2.0\,keV; 
\citealt{bt97,ken+02}). 

This seemingly `normal' pulsar, along with several tens of others, has been 
selected as one likely associated with the Galactic very-high energy 
(VHE; $>$100\,GeV) TeV sources, namely \hawc\ reported by the High-Altitude 
Water 
Cherenkov (HAWC) Observatory in the third HAWC catalog (3HWC; \citealt{3hawc})
and \lasso\ by the Large High Altitude Air Shower Observatory 
(LHAASO; \citealt{lhaaso19}) in the The First LHAASO Catalog of Gamma-Ray 
Sources \citep{cao+23}. The reasons for finding associations between VHE
sources and pulsars are the following. First, pulsars with $\tau_c\leq 100$\,kyr
can have pulsar wind nebulae (PWNe), which are considered as one primary type of
the Galactic TeV sources (e.g., \citealt{hess18,hesspwn18}). Second, as inspired
by the detection of the extended TeV emissions around two nearby 
pulsars, Geminga and Monogem \citep{abe+17}, a new type of TeV sources,
the so-called TeV halos that are powered by middle-aged pulsars, has been 
proposed \citep{lin+17,lb18}. Third, among more than 100 sources detected
in recent years with the VHE facilities, mainly the High Energy 
Spectroscopy System (HESS; \citealt{hess18}), the HAWC, and the LHAASO, 
a significant number of the sources
do not have typical known counterparts, i.e., the PWNe, the supernova
remnants (SNRs), or other types of VHE emitters \citep{hess18,3hawc,cao+23}. 
\begin{figure*}
\centering
\includegraphics[width=0.31\textwidth]{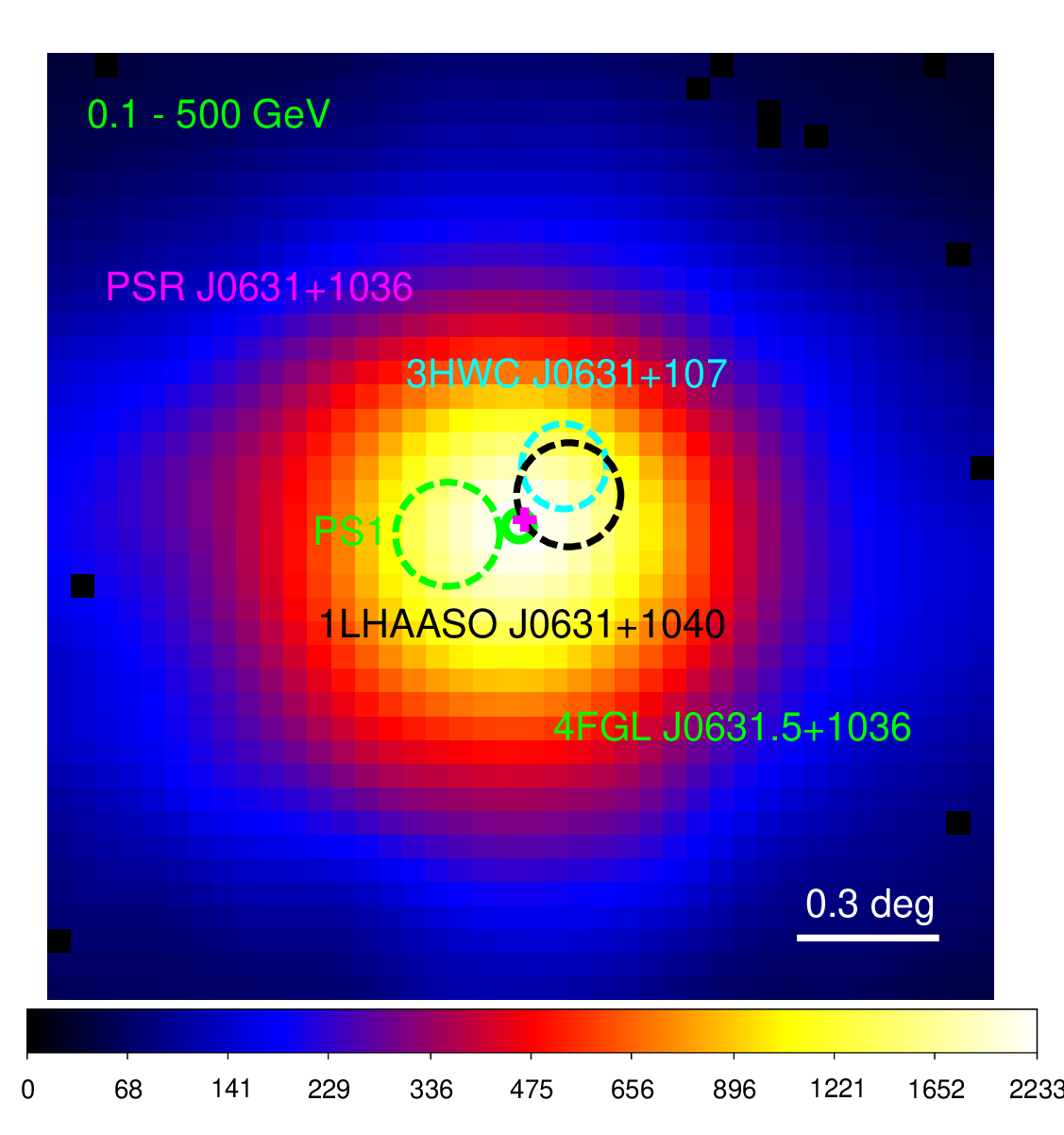}
\includegraphics[width=0.31\textwidth]{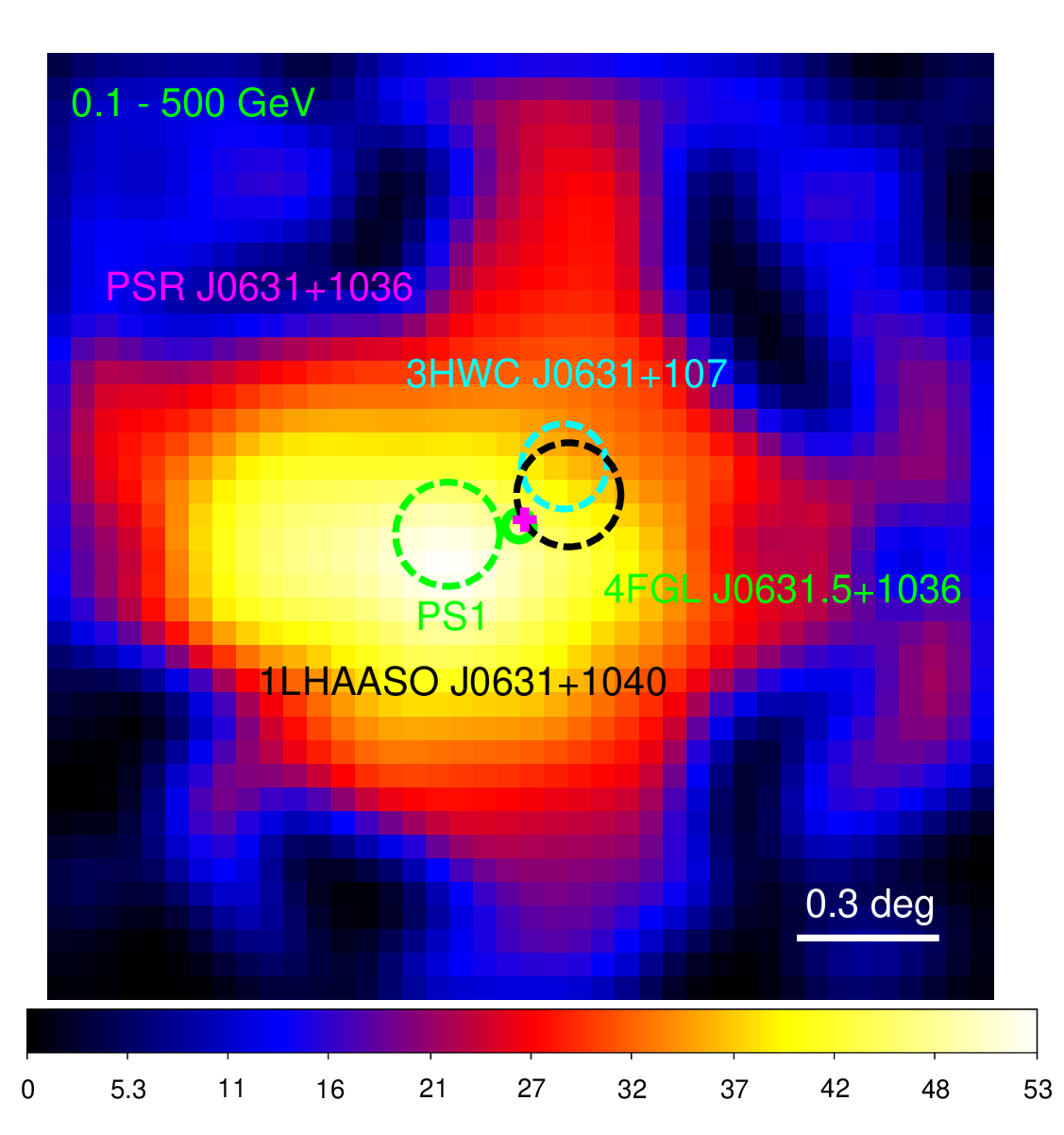}
\includegraphics[width=0.31\textwidth]{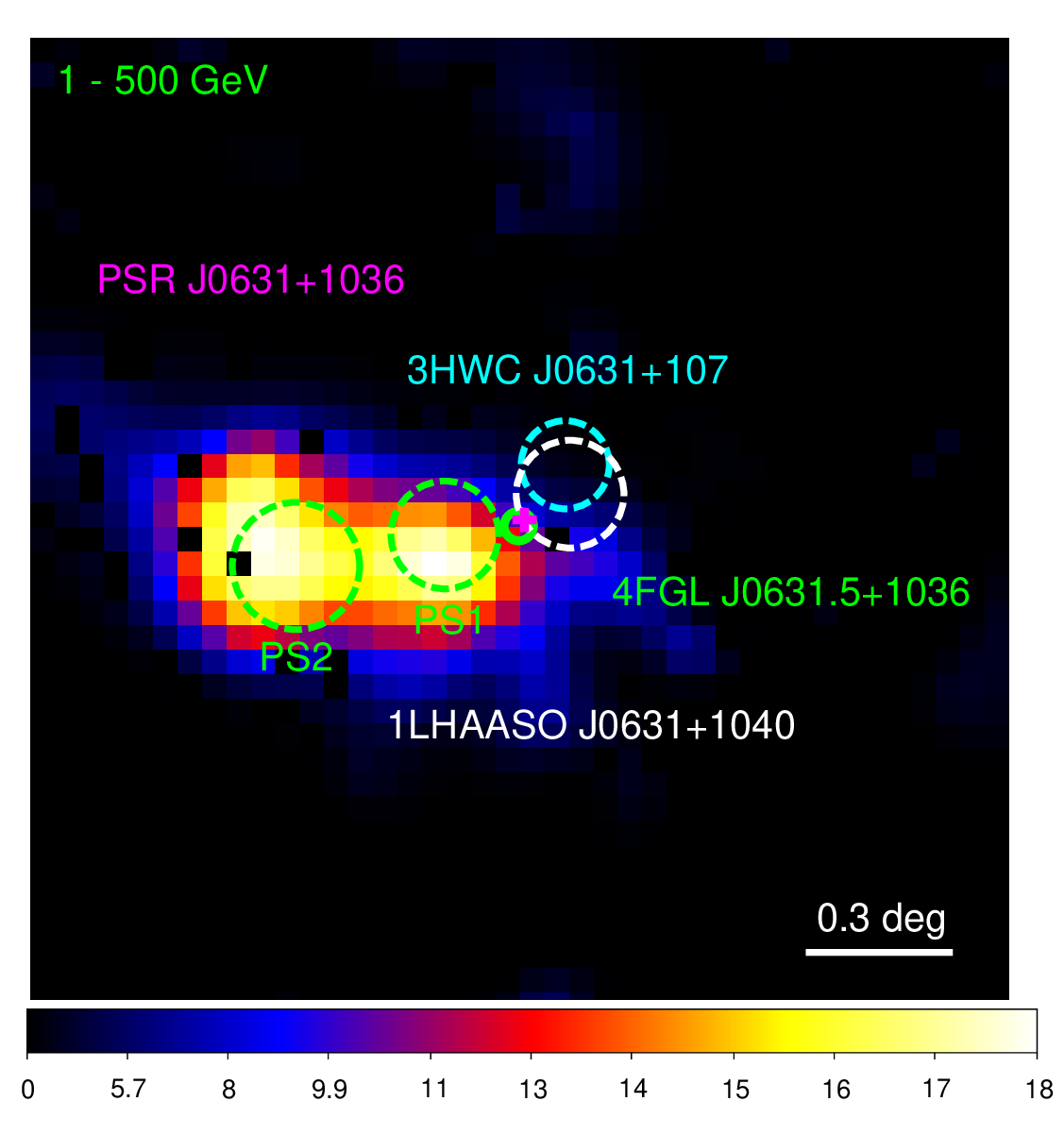}
\caption{TS maps for the region of PSR~\psr\ from the whole data 
	({\it left}, in 0.1--500\,GeV) and the pulsar's offpulse data 
	({\it middle}, in 0.1--500\,GeV; {\it right}, in 1--500\,GeV). The 
	pulsar's position (pink plus, with the green circle marking
	its LAT position) is within the 2$\sigma$
	HAWC error circle (the cyan dashed one being 1$\sigma$)
	and the LHAASO 95\% error circle (black or white dashed one). Two PSs,
	PS1 and PS2 (green dashed circles), are marked.}
\label{fig:tsmap}
\end{figure*}

Given these and intrigued by the third one described above, we have
been carrying out multi-wavelength studies of the TeV sources that
do not have obvious counterparts at other energy bands \citep{xin+22,zhe+23}. 
In our studies, we noted that 3HWC~\hawc\ (hereafter \hawc),
also LHAASO~\lasso\ given the positional match between these two 
sources, has a clean field in high energies. There are no known PWNe or 
SNRs found such as in the TeV online catalog 
(TeVCat; \citealt{wh08}), and PSR~\psr\ 
is the only notable source based on the position matches
(e.g., \citealt{cao+23}). Interestingly, this pulsar has bright GeV \gr\ 
emission, detected with the Large Area Telescope (LAT) onboard 
{\it the Fermi Gamma-ray Space Telescope (Fermi)} from the early observations
\citep{wel+10}. We thus conducted detailed analysis of the \fermi\ LAT data
for the pulsar. The analysis and results are described below in
Section~\ref{sec:obs}, based on which we argue that 
\hawc\ is likely a TeV halo powered by PSR~\psr; 
the related discussion is presented in Section~\ref{sec:dis}.

\section{Data Analysis}
\label{sec:obs}

\subsection{LAT data and source model}
We selected the 0.1--500\,GeV LAT data in the time range of
from 2008 Aug. 04 15:43:36 (UTC) to 2023 
Feb. 16 00:00:00 (UTC) from the latest \fermi\ Pass 8 database.
The region of interest (RoI) was 
$15^{\circ} \times 15^{\circ}$, centered at PSR~\psr.
As recommended by the LAT team\footnote{\footnotesize http://fermi.gsfc.nasa.gov/ssc/data/analysis/scitools/}, 
the events with quality flags of `bad' and zenith angles $\geq$ 90\arcdeg\ 
were excluded. We used the latest \fermi\ LAT 12-year source catalog 
(4FGL-DR3; \citealt{4fgl-dr3}) to construct a source model.
The sources within 15\arcdeg\ of the pulsar in the catalog 
were included in the source model, and their catalog spectral forms were
used. Also the background Galactic and extragalactic diffuse 
spectral models were included, with the files gll\_iem\_v07.fits and 
iso\_P8R3\_SOURCE\_V3\_v1.txt, respectively, used.

\subsection{Timing analysis of PSR~\psr}
\label{sec:ta}

PSR~\psr\ is bright in the LAT energy band and located in a clean field, as
shown in the left panel of Figure~\ref{fig:tsmap}, a test statistic (TS) map 
calculated for the source region from the whole LAT data (Section~\ref{sec:wo}).
The pulsar is the only 4FGL-DR3 source in the
$2^{\circ}\times 2^{\circ}$ TS map. Also seen is its positional match
to \hawc.

In order to check if there are other sources hiding in the bright emission
of the pulsar, we worked to obtain its pulsed emission through timing
analysis. On the first try,  we folded the photons
within a 6\arcdeg\ radius ($\sim$size of the
point spread function of LAT at 100\,MeV\footnote{https://www.slac.stanford.edu/exp/glast/groups/canda/lat\_Performance.htm}) 
aperture centered at the pulsar according to the ephemeris given in 
the LAT Gamma-ray Pulsar Timing Models 
Database\footnote{https://confluence.slac.stanford.edu/display/GLAMCOG/LAT+Gamma-ray+Pulsar+Timing+Models} \citep{ray+11}, 
but no clear pulse profile over the $\simeq$14.5\,yr long data could be 
obtained.

We then changed to use the method fully described in \citet{xin+22}.
In essence, we divided the data into sets of 200\,d, and assigned pulse phases 
to the photons according to the ephemeris in the Database \citep{ray+11} by
using the \fermi\ TEMPO2 plugin \citep{edw2006,hob2006}. 
We were able to obtain empirical Fourier template profiles before and after 
MJD~56770, generate the times of arrival (TOAs) for each set of $\sim 200$\,d 
data, and obtain timing solutions by fitting the TOAs with high-order 
frequency derivatives.
We could not extended the timing solutions to times longer than 
MJD~57930, probably due to the glitches of the pulsar at 
MJD~58341 \& 58352 \citep{low+21,bas+22}.
\begin{figure}
\centering
\includegraphics[width=0.47\textwidth]{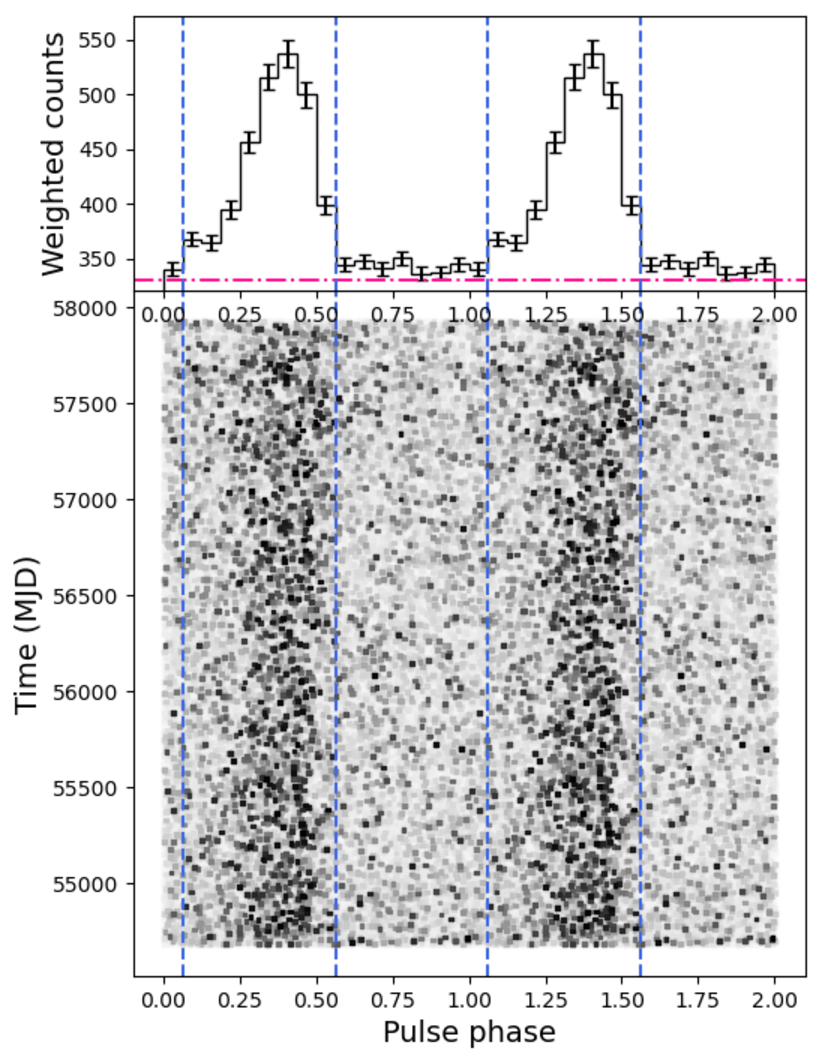}
	\caption{Phase-connected pulse profile ({\it top}) and two-dimensional
	phaseogram ({\it bottom}) of \psr\ during MJD 54682--57930.
	Two spin cycles are shown for clarity. The onpulse and offpulse
	phase ranges are marked by dashed lines.}
\label{fig:pha}
\end{figure}

With the two timing solutions,
the photons during the two time 
periods were folded respectively. The two pulse profiles had a phase mismatch 
of $\simeq$0.3075 (which was directly read off from the profiles
because of the clear pulse shape). After applying the phase shift to 
the photons of the second time period, the pulse profile 
over nearly 9\,yr was obtained (Figure~\ref{fig:pha}).
Based on the pulse profile, we defined phase 0.0625--0.5625 as the onpulse 
phase range and phases 0.0--0.0625 and 0.5625--1.0 as the offpulse phase 
ranges.
\begin{figure}
\centering
\includegraphics[width=0.47\textwidth]{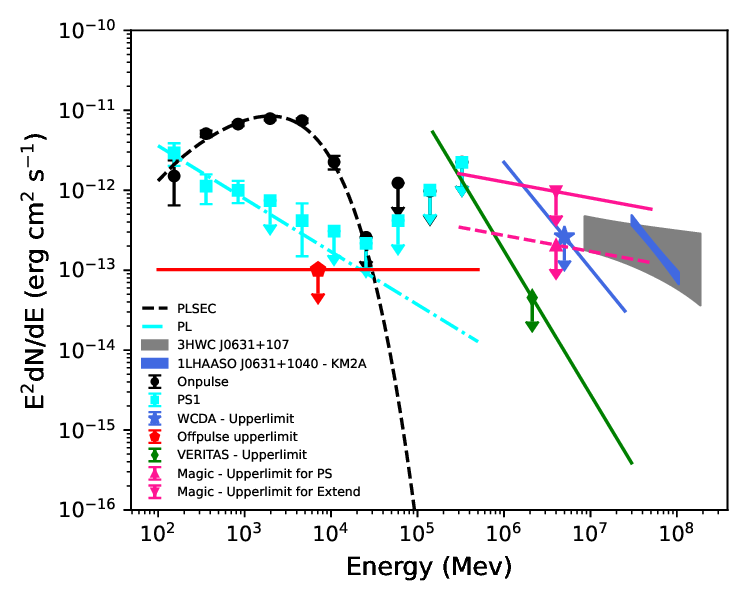}
\caption{\gr\ spectra of PSR~\psr\ in its onpulse data and PS1 in the offpulse
data, with their respective best-fit spectral models also shown. The red long
	bar indicates the flux upper limit in 0.1--500\,GeV at 
	the pulsar's position derived from the offpulse data. The HAWC and 
	LHAASO spectral measurements of \hawc\ are shown as the grey and blue
	shaded regions respectively. Other upper limits shown are on the pulsar
	(green line) from \citet{arc+19}, on the PWN (two pink lines) from 
	\citet{fer+17}, and in 1--25\,TeV on \hawc\ (blue line) 
	from \citet{cao+23}. 
}
\label{fig:spec}
\end{figure}

\subsection{Likelihood and spectral analysis}
\label{sec:la}

\subsubsection{Whole and onpulse data}
\label{sec:wo}

We performed standard binned likelihood analysis to the 0.1--500 GeV LAT data 
during the whole $\sim$14.5\,yr time period and the onpulse phase range during
the $\sim$9\,yr time period. The spectral parameters of
the sources within 5\arcdeg\ from the pulsar in the source model were set free,
while the spectral parameters of the other sources were fixed at the values 
given in 4FGL-DR3. In addition, the normalizations of the two background 
components were set free.

For the emission at the pulsar's position, in the whole data or
those of the onpules phase, we used a sub-exponentially cutoff power-law 
model (PLSEC; \citealt{4fgl20}), 
$\frac{dN}{dE} = N_{0} (\frac{E}{E_{0}})^{-\Gamma - \frac{d}{2} \ln(\frac{E}{E_{0}}) - \frac{db}{6} \ln^{2}(\frac{E}{E_{0}}) - \frac{db^{2}}{24} \ln^{3}(\frac{E}{E_{0}})}$, 
where $\Gamma$ and $d$ are the photon index and the local curvature at $E_{0}$ 
respectively, and $b$ is a measure of the shape of the exponential cutoff. 
We fixed $b=2/3$, which is such set for most of the $\gamma$-ray pulsars in
the LAT catalogs (e.g., \citealt{4fgl20,4fgl-dr3}).

The likelihood analysis results, including the TS values, are provided in 
Table~\ref{tab:ts}.  The parameter values of the pulsar are consistent with 
those given in 4FGL-DR3.
A 0.1--500\,GeV TS map of a $2^{\circ} \times 2^{\circ}$ region centered at 
the pulsar was calculated and is shown in the left panel of 
Figure~\ref{fig:tsmap}.

We extracted the \gr\ spectrum of PSR~\psr\ in the onpulse 
phase data. The spectral bands were set as 10 evenly divided in logarithm 
from 0.1 to 500\,GeV. In the analysis of obtaining fluxes in the bands,
the spectral normalizations of the sources within 5\arcdeg\ 
of the pulsar were set as free parameters, while all the other parameters 
of the sources were fixed at the values obtained from the above binned
likelihood analysis. For the spectral data points, we kept those with TS$\geq$4
and derived 95\% flux upper limits 
otherwise. The obtained spectrum is shown in Figure~\ref{fig:spec}.

\subsubsection{Offpulse data}
\label{sec:off}

With the offpulse phase ranges obtained above (Figure~\ref{fig:pha}), we 
examined the source region by first calculating a TS map in 0.1--500\,GeV
from the offpulse data. No source emission could be detected at the pulsar's
position, with a 95\% flux upper limit of 
$\sim$10$^{-13}$\,erg\,cm$^{-2}$\,s$^{-1}$ (assuming $\Gamma=2$ in 
0.1--500\,GeV; cf., Figure~\ref{fig:spec}). However residual emission 
(TS$\sim$50) southeast of 
the pulsar is seen (middle panel of Figure~\ref{fig:tsmap}). 

To further check the residual emission, a TS map in 1--500\,GeV was also 
obtained (right panel of
Figure~\ref{fig:tsmap}). As can be seen, there seemingly are two sources.
We ran \emph{gtfindsrc} in Fermitools to determine their positions. 
The obtained best-fit positions are (R.A., Decl.) = (98\fdg03, 10\fdg59) and 
(R.A., Decl.) = (98\fdg34, 10\fdg52) for point source 1 (PS1) and 2 (PS2), 
respectively, with the 1$\sigma$ nominal uncertainties of $0\fdg11$ 
and $0\fdg13$ (indicated in Figure~\ref{fig:tsmap}). By including PS1 or
PS1+PS2 in the source model, we performed the likelihood analysis to the
offpulse data. We found that PS2 was not significantly detected, indicated
by the likelihood value being similar to that when only PS1 was in 
the source model
(see Table~\ref{tab:ts}). We extracted the spectrum of PS1 
(Figure~\ref{fig:spec}), which could be fitted with a power law with
$\Gamma\simeq 2.66$).

\section{Discussion}
\label{sec:dis}

We conducted analysis of the LAT data for PSR~\psr, because of its possible
association with the TeV source \hawc\ and the absence of PWN/SNR-like 
counterparts in the source region.
By timing the pulsar, we were able to removed its pulsed emission in a 
$\sim$9\,yr time period of the data. No offpulse emission was detected
at the pulsar's position. Residual emission, PS1, was seen 
approximately 0\fdg16 east of the pulsar. The emission
was soft, mostly detectable at $\lesssim 1$\,GeV (Figure~\ref{fig:spec}). 
We checked the SIMBAD
database for possible counterparts to it, but no obvious ones (particularly
in radio or X-rays) were found
within its error circle. The nature of PS1 is thus not clear. Given the
positional offset and its soft emission, it is not likely in association 
with the pulsar or the TeV source.
\begin{figure}
\centering
\includegraphics[width=0.47\textwidth]{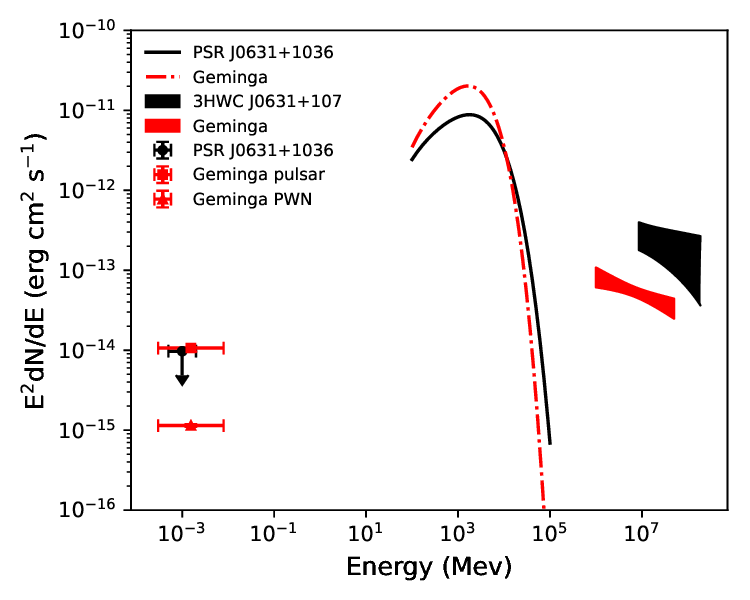}
\caption{Comparison of PSR~\psr\ with Geminga, where the X-ray flux upper limit
on the former and measurements of the latter and its PWN, \gr\ pulsar fluxes,
and TeV flux measurements of \hawc\ and the TeV halo of Geminga are shown.
The fluxes of Geminga are scaled to the distance of PSR~\psr.
}
\label{fig:comp}
\end{figure}

Then it is straightforward to note the similarities of PSR~\psr\ to Geminga.
They have similar $P$ values and are both \gr\ bright, while the former's 
$\tau_c$ is younger by a factor of $\sim$8 and $\dot{E}$ higher by
a factor of $\sim$5. Given these and our analysis results for the field,
we thus argue that \hawc\ is likely the TeV halo of PSR~\psr.
In Figure~\ref{fig:comp}, we compare this pulsar to Geminga. 
The latter's X-ray, \gr, and TeV halo fluxes, shown in the figure,
are scaled to the distance (2.1\,kpc) of the former,
where the nominal distance $250$\,pc is used for Geminga \citep{man+05}. 
As can be seen, the X-ray upper limit
on PSR~\psr\ \citep{ken+02} is approximately consistent with the X-ray 
fluxes of Geminga 
or its PWN \citep{pos+17}, where the fluxes of all the components of the 
latter's PWN
are added together as the flux in 0.3--8\,keV. Because the TeV halo of 
Geminga is extended with fine structures \citep{abe+17}, we adopt the 
flux measurement from the second HAWC catalog \citep{2hawc}, in which 
a 2\arcdeg\ extension was used. The scaled flux at 7\,TeV is 
$\sim$7$\times 10^{-16}$\,TeV$^{-1}$\,cm$^{-1}$\,s$^{-1}$, approximately 
$\sim$1/6 
of that of \hawc. Interestingly, the ratio is similar to that in $\dot{E}$
of Geminga to PSR~\psr. The size of the TeV halo of PSR~\psr\ would be 
roughly 0\fdg24 by taking that of Geminga as a standard \citep{lin+17}, 
smaller than the upper limit of 0\fdg30 set by the LHAASO \citep{cao+23}.

We further note that the emission of \hawc\ is hard, as LHAASO detected it 
in 25--100\,TeV
with $\Gamma\simeq 3.3$ but did not detect it in 1--25\,TeV 
(Figure~\ref{fig:spec}).
This spectral feature is similar to that of Geminga's TeV halo, indicated by
the LHAASO detection of $\Gamma\simeq 1.7$
in 1--25\,TeV and $\Gamma\simeq 3.7$ in $\geq$25\,TeV (i.e., the spectrum
likely peaks around $\sim$25\,TeV). This type of spectra is harder than 
those of PWNe, since the latter have $\Gamma\gtrsim 2$ in 1--10\,TeV and
thus some of them can be detected with \fermi\ LAT (\citealt{hesspwn18} and
references therein); indeed, part of the purpose of this work was to search
for a PWN in the offpulse data.
Hopefully with more data collected with LHAASO
in the near future, the similarity of the spectrum of \hawc\ to that of
the Geminga's TeV halo can be established, and thus firmly confirm the TeV
halo nature of \hawc.
\begin{table*}
	\centering
        \begin{center}
        \caption{Binned likelihood analysis results}
        \label{tab:ts}
        \begin{tabular}{lccccc}
        \hline
        \hline
        Phase Range (src)& $F_{0.1-500}/10^{-8} $   & $\Gamma$    & $d$ & TS & $\log(L)$  \\
                                      & (photons s$^{-1}$ cm$^{-2}$)  &                        &                   &      &                                               \\
        \hline
		Catalog values   & ---  & 1.85$\pm$0.05 & $0.54\pm 0.07$ & 2000  &   --- \\
        Whole data       & $3.01 \pm 0.23$     &$2.02 \pm 0.03$  & $ 0.47 \pm 0.05$ & 2570 & ---  \\
        \hline
        Onpulse     & $2.23 \pm 0.19$    &$1.96 \pm 0.04$ & $0.62 \pm 0.07$  & 2279 & --- \\
        \hline
        Offpulse (PS1)   & $1.35 \pm 0.29$  & $2.66 \pm 0.13$  & ---  &  60   & 763784.9 \\
		Offpulse (2PS-PS1) & $0.92 \pm 0.41$ & $2.63 \pm 0.17$ & --- & 29 & 763786.9 \\ 
		\ \ \ \ \ \ (2PS-PS2) & 0.45$\pm$0.36 & 2.53$\pm$0.25 & --- & 9 & ---\\
		\hline
\end{tabular}
\end{center}
\end{table*}

\begin{acknowledgements}
This research has made use of the SIMBAD database, operated at CDS, 
Strasbourg, France. This research is supported by 
the Basic Research Program of Yunnan Province
(No. 202201AS070005), the National Natural Science Foundation of
China (12273033), and the Original
Innovation Program of the Chinese Academy of Sciences (E085021002).
\end{acknowledgements}

\bibliographystyle{aasjournal}
\bibliography{thalo}

\end{document}